\documentstyle[seceq,epsf,preprint]{ptptex}

\newcommand{\bra}[1]{\langle {#1} |}     
\newcommand{\ket}[1]{| {#1} \rangle}     

\title{
Amplification of Quantum Meson Modes\\
in the Late Time of 
Chiral Phase Transition}

\author{
Kazu {\sc Watanabe},$^{1}$ 
Yasuhiko {\sc Tsue}$^{2}$ and Seiya {\sc Nishiyama}$^{2}$
}

\inst{
$^1$Department of Applied Science, Kochi University, Kochi 780-8520, Japan\\
$^{2}$Physics Division, Faculty of Science, Kochi University, Kochi 780-8520, 
Japan
}

\recdate{
\today
}

\abst{
It is shown that there exists a possibility of amplification of amplitudes of 
quantum pion modes with low momenta in the 
late time of chiral phase transition by using the Gaussian wave functional 
approximation in the O(4) linear sigma model. 
It is also shown that the amplification occurs in the mechanism 
of the resonance by forced oscillation as well as the parametric resonance 
induced by the small oscillation of the chiral condensate. 
These mechanisms are investigated in both the case of spatially homogeneous 
system and the spatially expanded system described by the Bjorken 
coordinate. 
}

\begin{document}

\maketitle

\section{Introduction}

One of the recent interests in the context of the 
relativistic heavy ion collisions is to clarify the nature of matter at 
very high energy density. 
Especially, it is interesting to 
investigate the dynamics of the chiral phase transition in connection with 
the problem of the formation of a disoriented chiral condensate (DCC)
\cite{ref:ANSELM,ref:BLAIZOT,ref:BJORKEN,ref:RAJAGOPAL,ref:AHW95,ref:BdV,ref:C,ref:SERREAU,ref:Salle,ref:Bet}.
Recently, Ikezi, Asakawa and one of the present authors (Y.T.) 
showed that there was a possibility of the formation of DCC 
by taking account of both the quantum fluctuations around the chiral 
condensate and the mode-mode coupling of quantum meson modes\cite{ref:IAT04}. 
In the formation of DCC, it is necessary that 
the low momentum components of the inhomogeneous chiral condensate were grown. 
Some authors pointed out that the parametric resonance was seen when 
the formation of DCC occurred\cite{ref:MM,ref:PR}.

Similarly, the amplitudes of quantum meson modes with low momenta 
may be grown in the chiral phase transition as well as the low momentum 
components of chiral condensate\cite{ref:TKI,ref:T}. 
In order to investigate the time evolution of the chiral condensate as an 
order parameter of chiral phase transition and of the fluctuation modes 
around it, it is suitable to adopt the time-dependent 
variational approach to dynamics of quantum fields in the 
O(4) linear sigma model in terms of 
a squeezed state or a Gaussian wave functional\cite{ref:VM,ref:TVM}. 
This method is based on the time-dependent variational method 
with a Gaussian wave functional by using a functional Schr\"odinger 
picture\cite{ref:JK79}. 

In this paper, we will show that the quantum meson modes are amplified  
even in the late time of chiral phase transition in both the cases 
of uniform and spatially expanding system in the O(4) linear sigma model. 
Then, it will be pointed out that the possible mechanism of amplification 
is the resonance mechanism by the forced oscillation as well as 
the parametric resonance which was seen in the DCC formation. 
The unstable region in the parameter space is also presented concretely. 
This paper is organized as follows. 
In the next section, we recapitulate the time-dependent variational 
approach with a Gaussian wave functional to the O(4) linear sigma model. 
The time evolution of chiral condensate and the quantum meson modes are 
governed by the Klein-Gordon type equation of motion and 
the Liouville-von Neumann type equation of motion, respectively. 
The numerical results are also given in \S 3. 
In \S 4, the time evolution of quantum meson modes in the late time 
of chiral phase transition is investigated. In both the case of the
uniform and the spatially expanding system, it is shown that 
there exists a possible solution of equation of motion for quantum meson
modes, which reveals 
a parametric resonance and/or resonance by forced oscillation induced 
by the small oscillation of the chiral condensate. 
In \S 5, the unstable regions are depicted for quantum meson modes, 
in which the amplitudes of meson modes are amplified by the mechanism 
of parametric resonance and/or the resonance of forced oscillation. 
The last section is devoted to the summary. 

\vspace{-0.3cm}

\section{Recapitulation of Gaussian wave functional approach 
to O(4) linear sigma model}

In this section, we derive the equations of motion for the chiral condensate 
and quantum meson modes in the O(4) linear sigma model. 
We apply the time-dependent variational method based on 
the functional Schr\"odinger picture to the O(4) linear sigma model. 
\subsection{Time-dependent variational approach with a Gaussian wave 
functional}

Following Refs.\citen{ref:VM} and \citen{ref:TVM}, 
we derive equations of motion for 
the chiral condensate and quantum fluctuations. 
We start with the following Hamiltonian :
\begin{eqnarray}\label{2-1}
\hat{H} = \int d^3{\mib x} \{ \frac{1}{2} \pi ^2_a({\mib x}) 
+ \frac{1}{2} (\nabla \varphi_a({\mib x}))^2 + \frac{m^2}{2} \varphi _a({\mib x})^2 
+ \frac{\lambda}{24} (\varphi _a({\mib x})^2)^2 - h \varphi _0({\mib x}) \} 
\ , \qquad\quad
\end{eqnarray}
where $a$ runs 0$\sim$3. The index 0 indicates the sigma field and 
1$\sim$3 indicate the pion fields.

We adopt the following Gaussian wave functional as a trial wave functional 
in the framework of the functional Schr\"odinger picture :
\begin{eqnarray}\label{2-2}
\Psi[\varphi({\mib x})]
={\cal N} \exp
\biggl(\!i\langle \bar{\pi}\ket{\varphi\!-\!{\bar \varphi}}
-\bra{\varphi\!-\!{\bar \varphi}}\!\left(\frac{1}{4G}-i\Sigma\right)\!
\ket{\varphi\!-\!{\bar \varphi}}\!\biggl) \ , \quad
\end{eqnarray}
where ${\cal N}$ is a normalization factor and 
$G_{ab}({\mib x},{\mib y},t)$, $\Sigma_{ab}({\mib x},{\mib y},t)$, 
$\bar{\varphi}_a({\mib x},t)$ and $\bar{\pi}_a({\mib x},t)$ 
define the real and 
imaginary parts of the kernel of the Gaussian and its average position and 
momentum. 
Here, we have used simple notations as 
\begin{eqnarray}\label{2-3}
& &\langle {\bar \pi} \ket{\varphi}=\int d^3{\mib x} 
\sum_{a=0}^3{\bar \pi}_a({\mib x},t)\varphi_a({\mib x}) \ , 
\nonumber\\
& &\bra{\varphi}G\ket{\varphi}\!=\!\int\!\!\!\!\int\!\!
 d^3{\mib x}d^3{\mib y}\!\!
\sum_{a,b=0}^3\!
\varphi_a({\mib x})G_{ab}({\mib x},{\mib y},t)\varphi_b({\mib x})
\ , \ \ 
\end{eqnarray}
and so on. 
In the functional Schr\"odinger picture, the operator $\pi_a({\mib x})$, 
which is a conjugate operator of the field operator $\varphi_a({\mib x})$, 
is regarded as $-i\delta/\delta\varphi_a({\mib x})$. 
The expectation value $\langle {\hat O} \rangle$ 
for the field operator ${\hat O}$ is easily calculated such as 
\begin{eqnarray}\label{2-4}
&&\langle \varphi_a({\mib x}) \rangle = \bar\varphi_a({\mib x},t) \ , 
\nonumber\\
&&\langle \pi_a({\mib x}) \rangle = \bar\pi_a({\mib x},t) \ , 
\nonumber\\
&&\langle \varphi_a({\mib x}) \varphi_b({\mib y})\rangle 
= \bar\varphi_a({\mib x},t) \bar\varphi_b({\mib y},t) + 
G_{ab}({\mib x},{\mib y},t) \ , \nonumber\\
&&\langle \pi_a({\mib x}) \pi_b({\mib y})\rangle 
= \bar\pi_a({\mib x},t) \bar\pi_b({\mib y},t)
+ \frac{1}{4} G_{ab}^{-1}({\mib x},{\mib y},t) 
+ 4\Sigma\cdot G\cdot \Sigma_{ab}({\mib x},{\mib y},t) \ , \qquad
\end{eqnarray}
where we have used the following shorthanded notation : 
\begin{eqnarray}
& &A\!\cdot\! B_{ab}({\mib x},{\mib y},t) 
=\!\sum_{c=0}^3\!\! \int d^3{\mib z}
A_{ac}({\mib x},{\mib z},t) B_{cb}({\mib z},{\mib y},t) 
\ . \nonumber
\end{eqnarray}
It is understood that $\bar{\varphi}$ represents the mean field which is 
identical with the chiral condensate in this model, and $G$ represents 
quantum fluctuations around the mean field.

The time dependence of the Gaussian wave functional is 
determined by the time dependence of variational functions 
$G$, $\Sigma$, 
$\bar{\varphi}$ and $\bar{\pi}$, 
which is governed by the time-dependent variational principal :
\begin{eqnarray}\label{2-5}
\delta \int dt 
\biggl\langle i \frac{\partial}{\partial t} - \hat{H} \biggl\rangle =0 \ . 
\end{eqnarray}
The variational equations give the canonical equations of motion 
for $(\bar{\varphi}_a,\bar{\pi}_a)$ and $(G_{ab},\Sigma_{ab})$ with 
the Hamiltonian $\langle {\hat H}\rangle$. 
As a result, we obtain the following equations of motion :
\begin{eqnarray}
&&\dot{\bar{\varphi}}
= \bar{\pi} \ , \nonumber\\
&&\dot{\bar{\pi}} 
= \left\{\! (\nabla ^2 - m^2 -\frac{\lambda}{6} {\bar \varphi}^2 
-\frac{\lambda}{6}{\rm tr}G({\mib x},{\mib x}))
-\frac{\lambda}{3}G({\mib x},{\mib x})\!\right\} \bar{\varphi}
 + h \delta_{a0} \ , 
 \label{2-6}\\
&&\dot{G} = 2(\Sigma\cdot G + G\cdot \Sigma) \ , 
\nonumber\\
&&\dot{\Sigma}=- \frac{1}{8}G^{-2} + 2\Sigma ^2
+\frac{1}{2}(-\nabla ^2 + m^2 +\frac{\lambda}{12} 
({\bar \varphi}^2+ {\rm tr} G)) 
+\frac{\lambda}{6}({\bar \varphi}{\bar \varphi} + G) \ . 
\label{2-7}
\end{eqnarray}
These are a set of basic equations of motion for the chiral condensate 
${\bar \varphi}$ and quantum fluctuations $G$.

\subsection{Reformulation of equations of motion for quantum fluctuations}

In this subsection, we reformulate the equations of motion for 
quantum fluctuations in order to investigate clearly the time evolution 
of quantum meson modes. 
This reformulation has been firstly carried out in Ref.\citen{ref:TVM}. 
We introduce the reduced density matrix ${\cal M}_{ab}$ :
\begin{eqnarray}\label{2-8}
{\cal M}_{ab}({\mib x},{\mib y},t)
+\frac{1}{2}\delta^3({\mib x}-{\mib y} )
= \left(
\begin{array}{@{\,}cc@{\,}}
-i\langle {\hat \varphi}_a({\mib x}){\hat \pi}_b({\mib y})\rangle 
 &
\langle {\hat \varphi}_a({\mib x}){\hat \varphi}_b({\mib y})\rangle \\
\langle {\hat \pi}_a({\mib x})
{\hat \pi}_b({\mib y})\rangle &
i\langle {\hat \pi}_a({\mib x}){\hat \varphi}_b({\mib y})\rangle 
\end{array}
\right) \ , 
\end{eqnarray}
where $\hat{\varphi}_a= \varphi_a - \bar{\varphi}_a$ and 
$\hat{\pi}_a= \pi_a - \bar{\pi}_a$. 
Thus, the reduced density matrix can be expressed in terms of $G$ and $\Sigma$ 
as 
\begin{eqnarray}\label{2-9}
&&{\cal M}=\left(
\begin{array}{@{\,}cc@{\,}}
-2iG\!\cdot\!\Sigma & G \\
\frac{1}{4}G^{-1} + 4\Sigma\!\cdot\! G\!\cdot\! \Sigma \ 
& \ 2i\Sigma\!\cdot\! G
\end{array}
\right) \ . 
\end{eqnarray}
The time dependence of this reduced density matrix is governed by the 
Liouville-von Neumann type equation :\cite{ref:VAUTHERIN} 
\begin{eqnarray}\label{2-10}
i\dot{\cal M}_{ab}({\mib x},{\mib y},t) =  [{\cal H},{\cal M}]_{ab}({\mib x},{\mib y},t) \ , 
\end{eqnarray}
where the generalized Hamiltonian ${\cal H}_{ab}$ has particularly 
simple form as 
\begin{eqnarray}\label{2-11}
&&{\cal H}_{ab}({\mib x},{\mib y},t) = 
\left(
\begin{array}{@{\,}cc@{\,}}
0 & \delta_{ab} \\
\Gamma_{ab}({\mib x},t) & 0
\end{array}\right) \delta^3({\mib x} - {\mib y}) 
\end{eqnarray}
with 
\begin{eqnarray}\label{Mass}
&&\Gamma_{ab}({\mib x},t) 
= -\nabla^2 \delta_{ab}+M_{ab}^2({\mib x},t) \ , \\
&&M_{ab}^2({\mib x},t)= \left(\!\!
m^2+ \frac{\lambda}{6}\bar\varphi^2({\mib x},t) + 
\frac{\lambda}{6}{\rm tr} G({\mib x},{\mib x},t) \!\!\right)\!\! 
\delta_{ab} \nonumber\\
&&\qquad\qquad\ \ 
+ \frac{\lambda}{3}\bar\varphi_a({\mib x},t) \bar\varphi_b({\mib x},t) 
+ \frac{\lambda}{3}G_{ab}({\mib x},{\mib x},t) \ .
\nonumber
\end{eqnarray}
Equations of motion derived from (\ref{2-10}) are equivalent to (\ref{2-7}).

From the structure of ${\cal M}$, 
it can be checked that the reduced density matrix satisfies 
a relation ${\cal M}^2=1/4$. 
The eigenvalues of the reduced density matrix are thus $\pm 1/2$. 
We introduce an eigenvector of ${\cal M}_{ab}$, $(u_a,v_a)$, 
with eigenvalue 1/2, i.e.,
\begin{eqnarray}\label{2-12}
\int\!\! d^3{\mib y}{\cal M}_{ab}({\mib x},{\mib y},t)
\left(
\begin{array}{@{\,}c@{\,}}
u_b({\mib y},t)\\
v_b({\mib y},t)
\end{array}\right) = 
\ +\frac{1}{2}
\left(
\begin{array}{@{\,}c@{\,}}
u_a({\mib x},t) \\
v_a({\mib x},t)
\end{array}\right) \ . 
\end{eqnarray}
Then, $(u^{*}_a,-v^*_a)$ is also an eigenvector with eigenvalue $-1/2$.  
The eigenvectors are conveniently normalized to
\begin{eqnarray}\label{normalization}
\sum_{a=0}^3\! \int\!\! d^3{\mib x} 
(u^{(n)}_a\!({\mib x},t)^*v^{(m)}_a\!({\mib x},t) 
+ v^{(n)}_a\!({\mib x},t)^*u^{(m)}_a\!({\mib x},t))=\pm \delta _{mn} 
\end{eqnarray}
with a $\pm$ sign for eigenvalues $\pm 1/2$. 
For the particular normalization condition we have adopted the eigenvectors 
$u_a$ and $v_a$ provide the following 
spectral decomposition for the reduced density matrix ${\cal M}_{ab}$:
\begin{eqnarray}\label{2-13}
{\cal M}_{ab}({\mib x},{\mib y},t) &=&
 \frac{1}{2} \sum_n \biggl[\left(\!
 \begin{array}{@{\,}c@{\,}}
	u_a^{(n)}({\mib x},t) \\
	v_a^{(n)}({\mib x},t)
 \end{array}\!\right)\!\!
 \left(\!
 \begin{array}{@{\,}cc@{\,}}
	v^{(n)}_b({\mib y},t)^* & u^{(n)}_b({\mib y},t)^*
	\end{array}\!\right)
	 \nonumber\\
& &\qquad\quad
+
\left(\!
\begin{array}{@{\,}c@{\,}}
	u^{(n)}_a({\mib x},t)^* \\
	-v^{(n)}_a({\mib x},t)^*\!\!
	\end{array}\!\right)\!\!
 \left(\!
 \begin{array}{@{\,}cc@{\,}}
	-v_b^{(n)}({\mib y},t) & u_b^{(n)}({\mib y},t)\!
	\end{array}\!\right) 
\biggl]  \ , \qquad
\end{eqnarray}
where sum runs over eigenstates with eigenvalues +1/2 only.
From Eqs.(\ref{2-9}) and (\ref{2-13}), the Gaussian kernel $G$ and $\Sigma$ 
can be expressed 
in terms of $u$ and $v$, for example, 
\begin{eqnarray}\label{G}
G_{ab}({\mib x},{\mib y},t)=
\frac{1}{2}\sum_n (u_a^{(n)}\!({\mib x},t)u_b^{(n)}\!({\mib y},t)^*\!
+u_a^{(n)}\!({\mib x},t)^*u_b^{(n)}\!({\mib y},t)) \ .\qquad
\end{eqnarray}
Thus, $u_a$ can be regarded as mode functions for quantum meson modes. 

The equations of motion for mode functions 
can also be written by using the generalized Hamiltonian matrix : 
\begin{eqnarray}\label{2-14}
i\partial_t
\left(\begin{array}{@{\,}c@{\,}}
	u_a({\mib x},t) \\
	v_a({\mib x},t)
\end{array}\right)
	=
	\left(
	\begin{array}{@{\,}cc@{\,}}
0 & \delta_{ab} \\
\Gamma_{ab}({\mib x},t) & 0 
\end{array}\right) 
\left(
\begin{array}{@{\,}c@{\,}}
	u_b({\mib x},t) \\
	v_b({\mib x},t)
	\end{array}\right) \ . 
\end{eqnarray}
The normalization condition (\ref{normalization}) 
is preserved by these equations.
Using the explicit form of the mean field operator $\Gamma_{ab}$ , we see that 
the mode function $u_a$ satisfy the set of coupled Klein-Gordon-type equations
\begin{eqnarray}\label{2-15}
\left( {\Box}\delta_{ab} + M^2_{ab}({\mib x},t) \right)u_b({\mib x},t) = 0 
\ . 
\end{eqnarray}
The matrix elements $M_{ab}^2$ are found to be diagonal in the isospin index, 
i.e.,$M^2_{ab} = M^2_a \delta_{ab}$. 
Let us assume that $\bar{\varphi}_0 \neq 0$ and 
$\bar{\varphi}_1=\bar{\varphi}_2=\bar{\varphi}_3=0$, that is, 
the chiral condensate points in the sigma direction. 
Then the Eqs.(\ref{2-6}) and (\ref{2-15}) read 
\begin{eqnarray}
& &\left\{ \Box + M^2_0({\mib x},t) 
- \frac{\lambda}{3} \bar{\varphi}_0({\mib x},t)^2 \right\}
\bar{\varphi}_0({\mib x},t) =h \ , 
\label{2-17}\\
& &\left( \Box + M^2_{a}({\mib x},t) \right) u_a({\mib x},t) = 0 \ . 
\label{2-18}
\end{eqnarray}

It should be noted here that, 
if the condensate is uniform spatially with translation invariance, namely, 
the condensate depends on only time, 
it is possible to carry out the Fourier transformation for 
$u_a({\mib x},t)$ as follows :
\begin{eqnarray}\label{2-19}
u_a({\mib x},t) = \frac{1}{\sqrt{(2\pi)^3}} \int d^3{\mib k} 
u_a^{\bf k}(t) 
e^{i {\bf k} \cdot {\bf x}} \ . 
\end{eqnarray}
Then, the above forms are substituted into Eq.(\ref{2-18}) and 
${\bar \varphi}_0$ depends only on time, 
we found following equations : 
\begin{eqnarray}
& &\left\{\partial_t^2 + M^2_0({\mib x},t) 
- \frac{\lambda}{3} \bar{\varphi}_0(t)^2 \right\}
\bar{\varphi}_0(t) =h \ , 
\label{2-20}\\
& &\left\{ \partial_t^2 
+ {\mib k}^2 + M^2_{a}({\mib x},t) \right\} 
u_a^{{\bf k}}(t) = 0 \ . 
\label{2-21}
\end{eqnarray}
These equations are investigated later.

\subsection{Geometry for the spatially expansion}
In this subsection, we derive the equations of motion 
that is taken account of the geometry of one-dimensional spatial expansion 
in $z$-direction.\cite{ref:BRAGHIN} 
For this purpose, the convenient variables, namely, 
the proper time $\tau$ and usual rapidity variable $\eta$ are defined by
\begin{eqnarray}\label{2-22}
\tau = \sqrt{t^2 - z^2} \ , \qquad 
\eta = \frac{1}{2}\ln{\frac{t+z}{t-z}} 
\end{eqnarray}
for one dimensional scaling case\cite{ref:BJORKEN83}. 
D'Alembertian in Eqs.(\ref{2-17}) and (\ref{2-18}) 
is rewritten by using these variables :
\begin{eqnarray}\label{2-23}
 \Box = \partial _{\tau}^2 + \frac{1}{\tau}\partial _{\tau} 
 - \frac{1}{\tau^2}\partial _{\eta}^2 - \partial _{\bot}^2 \ , 
\end{eqnarray}
where $\partial_{\bot}^2 = \partial_x^2 + \partial_y^2$.

If the condensate does not depend on 
$({\mib x}_{\bot},\eta)$, namely, 
the condensate depends on only proper time $\tau$, 
the quantum fluctuation modes $u_a$ can be expressed as follows \cite{ref:C}:\break
\begin{eqnarray}\label{2-26}
& &u_a({\mib x}_{\bot},\eta,\tau) = \frac{1}{\sqrt{\tau}}
\!\int_k\!\! u_a^{[{\bf k}]}(\tau)
e^{i k_{\eta} \eta} e^{i {\bf k}_{\bot} \cdot {\bf x}_{\bot}} \ ,
\end{eqnarray}
where we have defined $\int_k=\int\!d^2{\mib k}_{\bot} dk_{\eta}/
(2\pi)^{3/2}$. 
Then, the equations of motion (\ref{2-17}) and (\ref{2-18}) 
are recast into the following equations : 
\begin{eqnarray}
& &\left[(\partial _{\tau}^2 + \frac{1}{\tau}\partial _{\tau}) + M_0^2(\tau) 
- \frac{\lambda}{3}{\bar \varphi}_0^2(\tau) \right]{\bar \varphi}_0(\tau) 
= h \ , 
\label{2-27}\\
& & \left[\! \partial _{\tau}^2  
 + \frac{k_{\eta}^2 + 1/4}{\tau^2}  + {\mib k}_{\bot}^2 
 + M_a^2({\mib x}_{\bot},\eta,\tau)\! \right]\!u_a^{[\bf k]}(\tau) = 0 \ ,
 \nonumber\\
& &
 \label{2-28}
\end{eqnarray}
where momentum $k_{\eta}$ appears in the combination 
$(k_{\eta}^2+1/4)/{\tau^2}$.

\section{Numerical calculation}

Let us demonstrate qualitatively 
the time evolution of the mean field, which is in the 
sigma-direction only, and of quantum meson mode functions in the case 
of an uniform system in Eqs. (\ref{2-20}) and (\ref{2-21}) 
without spatially expansion. 
We assume $\bar{\varphi}_i = 0$ with $i = 1\sim 3$.
In the numerical calculation, we adopt the box normalization with the spatial 
length being $L$ in each direction. 
We then impose the periodic boundary conditions for the fluctuation modes, 
namely the allowed values of momenta are $k_x = (2 \pi/L)n_x$ and so on, 
where $n_x$ is integer.
The fluctuation modes labeled by $(n_x,n_y,n_z)$ are included each direction 
up to 
$n^2\equiv n_x^2+n_y^2+n_z^2 = 8^2$.
This corresponds to the momentum cutoff $\Lambda \sim 1{\rm GeV}$(990 MeV) 
since we have adopted 
the collisional region as $L^3 = (10\  {\rm fm})^3$.
Further, we assume that fluctuation modes of the pion fields 
are identical one another, which are denoted as 
$u_1^{\bf k} = u_2^{\bf k} = u_3^{\bf k} \equiv u_{\pi}^{\bf k}$. 

\begin{figure}[htb]
 \parbox{\halftext}
 {  \epsfxsize=5.5cm  
  \centerline{\epsfbox{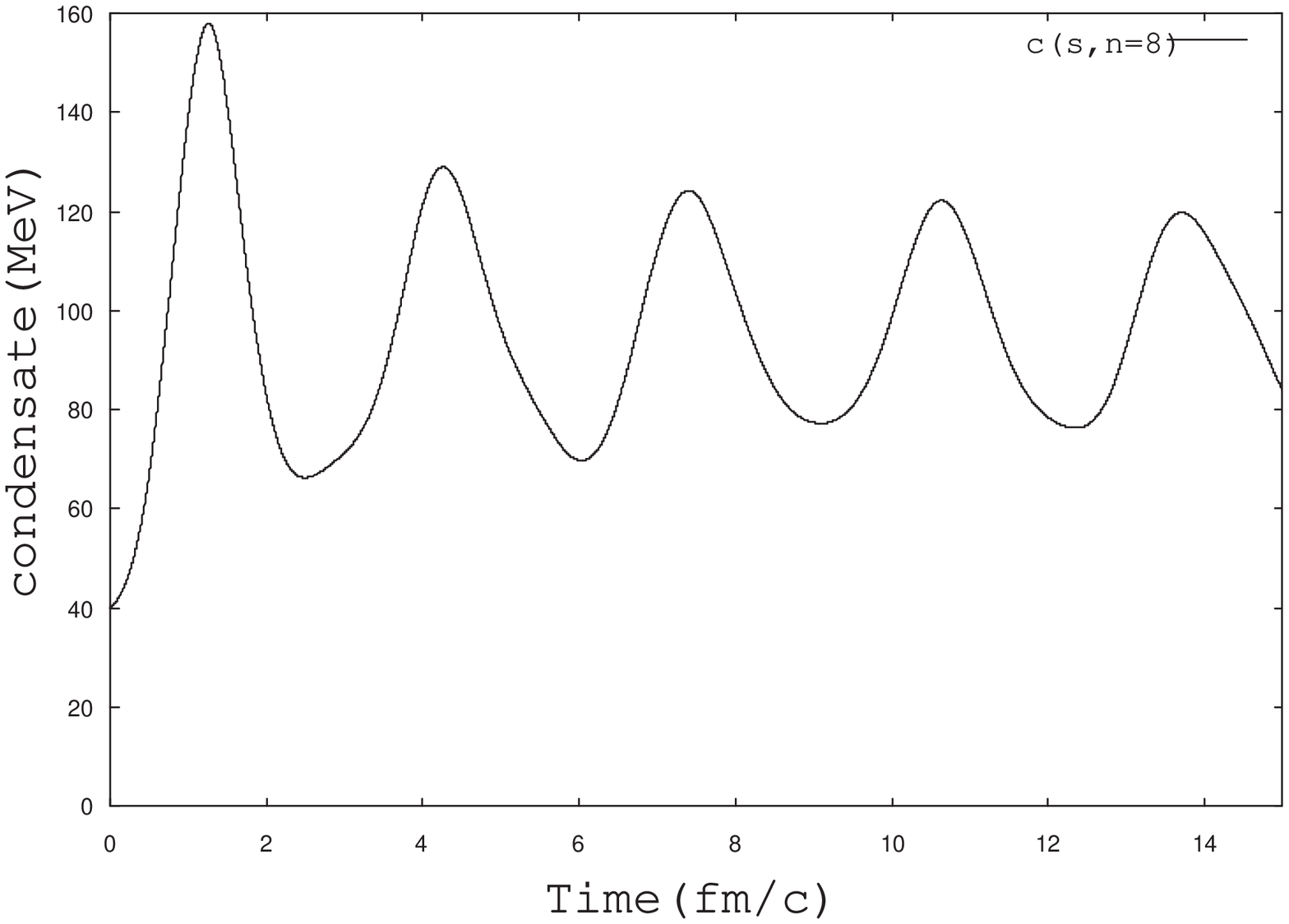}}
  		\caption{The time evolution of the chiral condensate is 
    		     depicted. The horizontal and vertical axes represent time 
    		     and the value of chiral condensation, respectively.}}
   \label{fig:cond1}
 \hspace{3mm}
 \parbox{\halftext}
 {  \epsfxsize=5.5cm  
  \centerline{\epsfbox{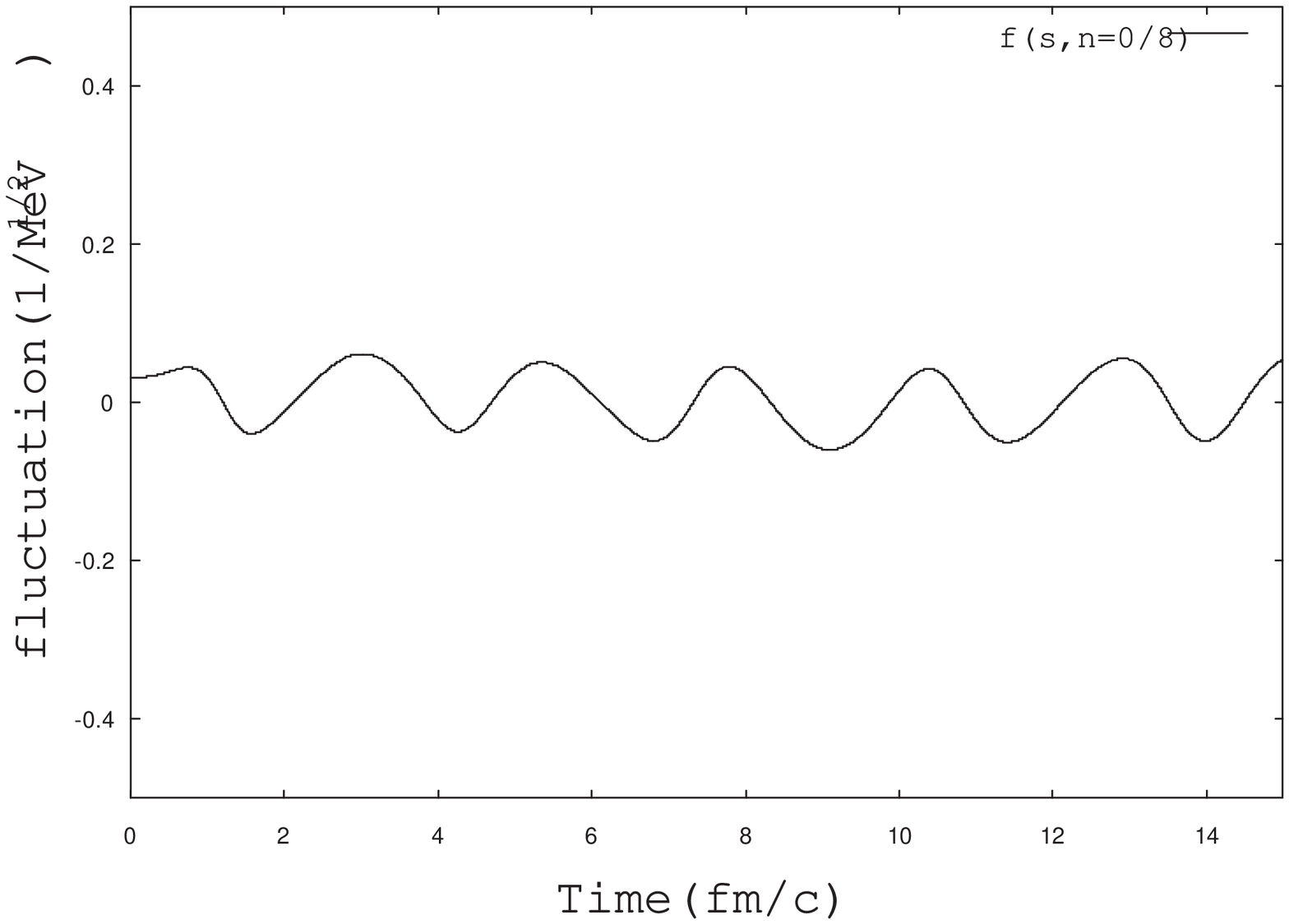}}
  		\caption{The time evolution of the fluctuation mode with $n=0$ 
                in the $\sigma$-direction is depicted.}}
   \label{fig:sigma0}
\end{figure}
\begin{figure}[htb]
 \parbox{\halftext}
 {  \epsfxsize=5.5cm  
  \centerline{\epsfbox{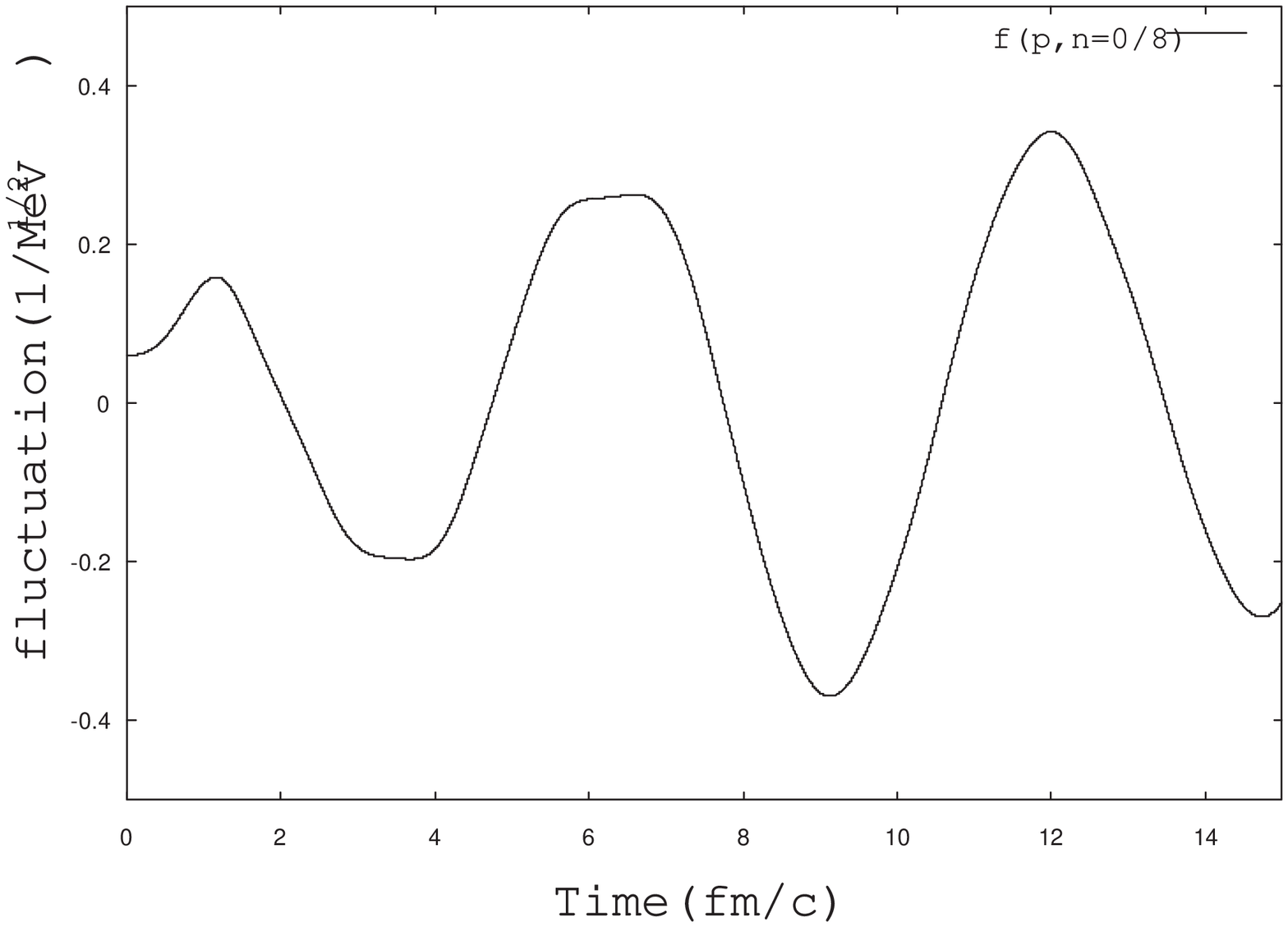}}
  \caption{The time evolution of the fluctuation mode with $n=0$ 
                in the $\pi$-direction is depicted.}}
   \label{fig:pi0}
 \hspace{3mm}
 \parbox{\halftext}
 {  \epsfxsize=5.5cm  
  \centerline{\epsfbox{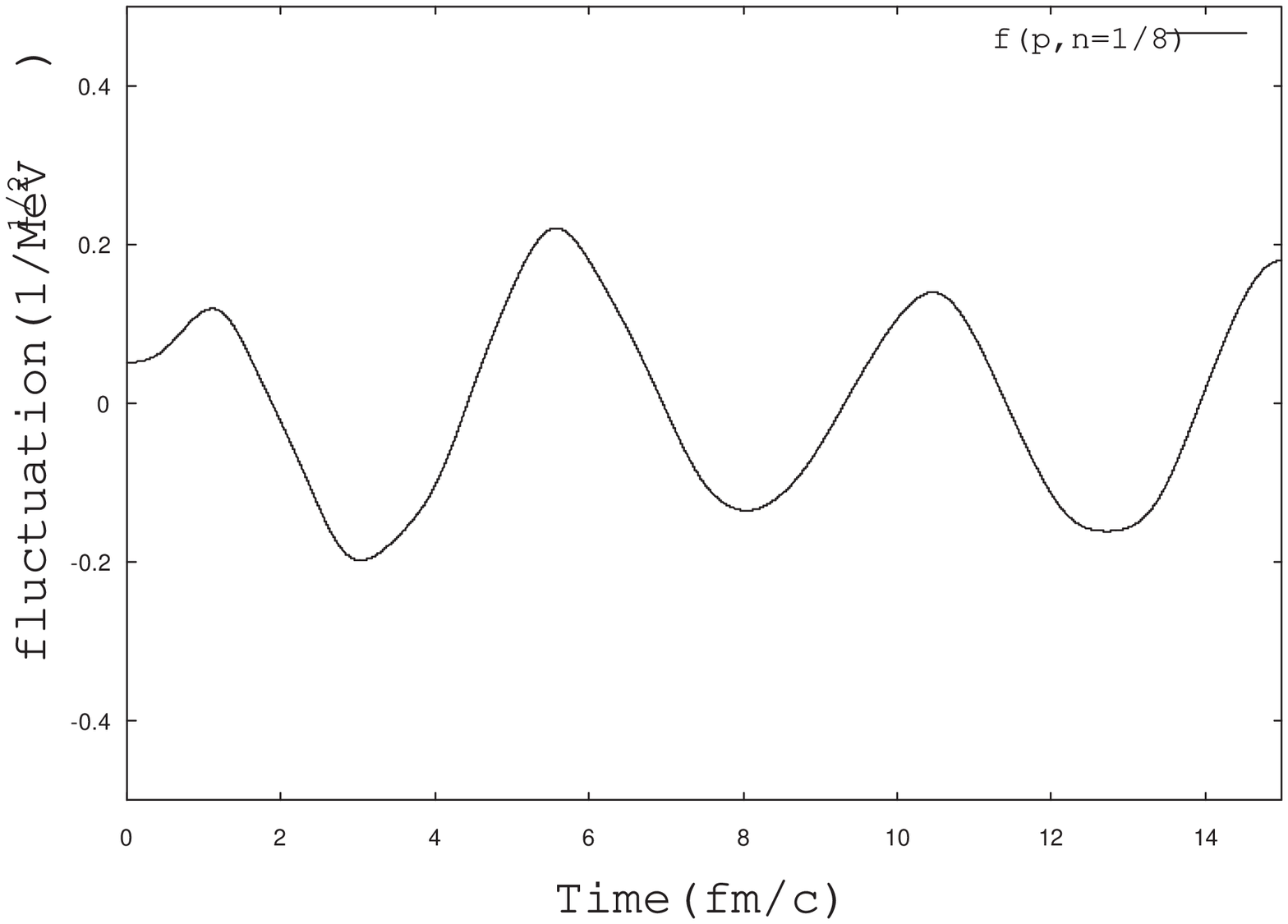}}
  \caption{The time evolution of the fluctuation mode with $n=1$ 
                in the $\pi$-direction is depicted.}}
   \label{fig:pi1}
\end{figure}
\begin{figure}[htb]
 \parbox{\halftext}
 {  \epsfxsize=5.5cm  
  \centerline{\epsfbox{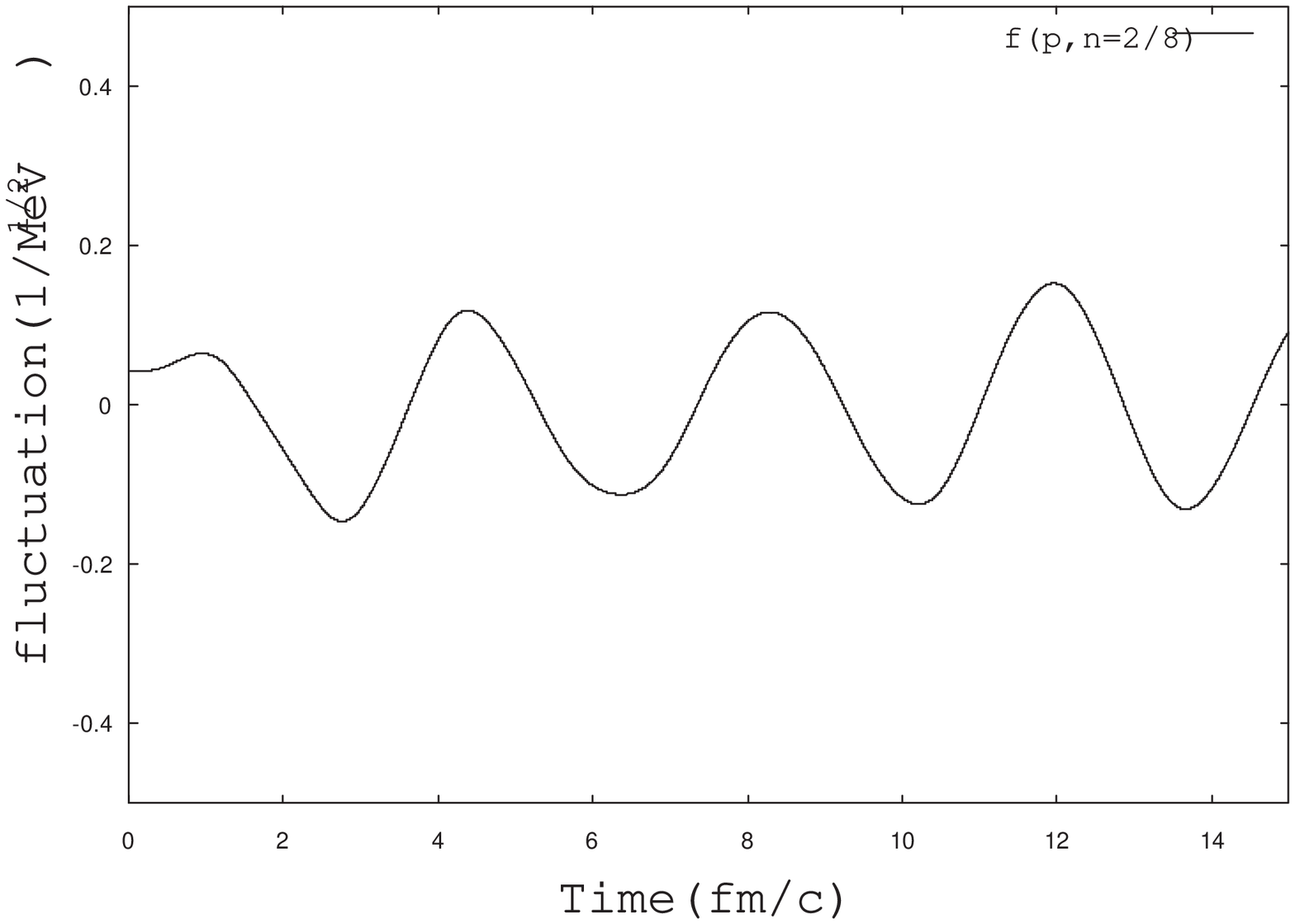}}
  \caption{The time evolution of the fluctuation mode with $n=2$ 
                in the $\pi$-direction is depicted.}}
   \label{fig:pi2}
 \hspace{3mm}
 \parbox{\halftext}
 {  \epsfxsize=5.5cm  
  \centerline{\epsfbox{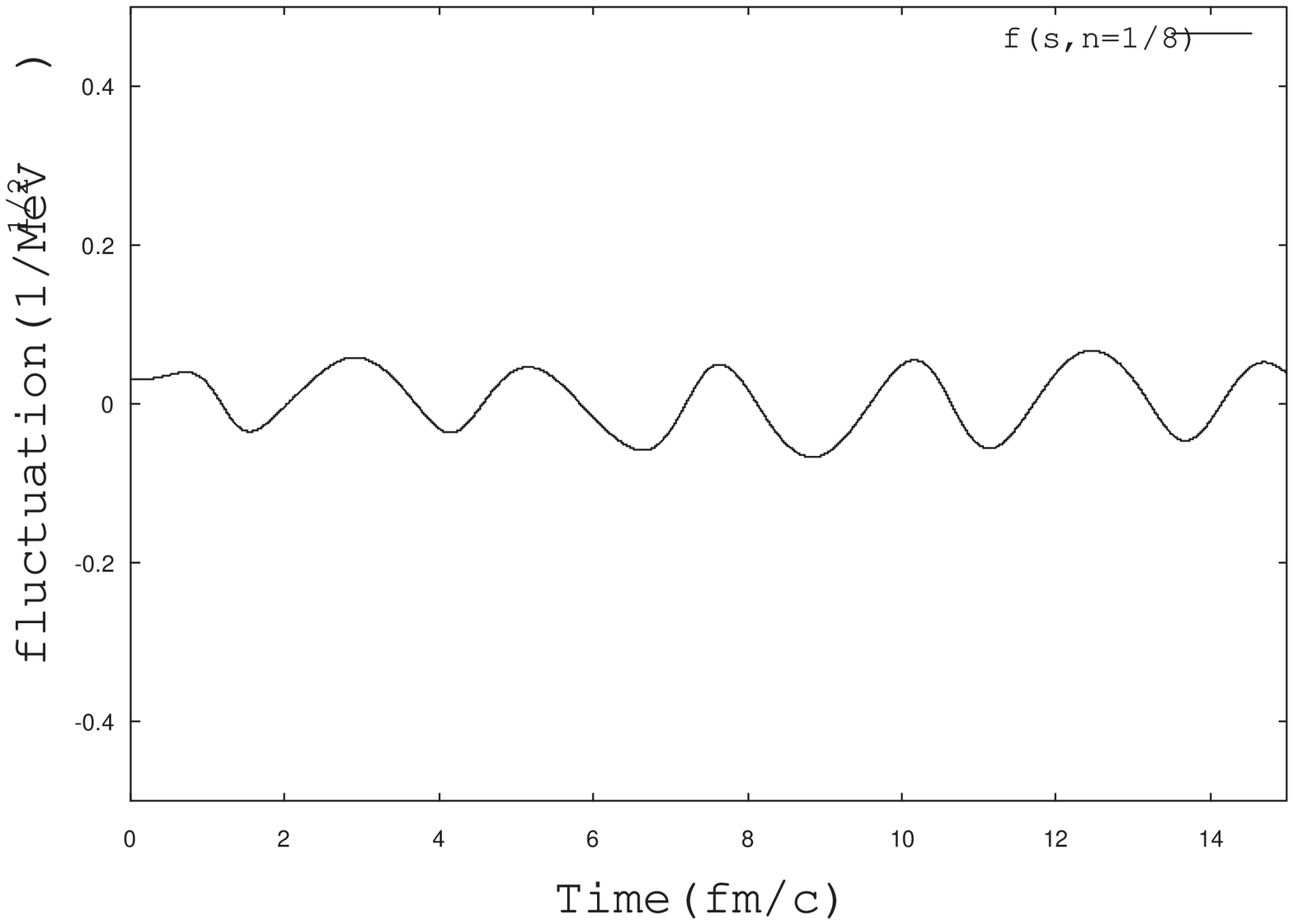}}
  \caption{The time evolution of the fluctuation mode with $n=1$ 
                in the $\sigma$-direction is depicted.}}
   \label{fig:sigma1}
\end{figure}

In Fig.1, the time evolution of chiral condensate is depicted. 
In order to avoid the complexity of problems with respect to the initial 
conditions in the relativistic heavy ion collisions, we only demonstrate 
the time evolution qualitatively with $\bar{\varphi}_0(t=0)=40$ MeV and 
$\dot{\bar{\varphi}}_0 = 0$. 
It is shown that the chiral condensate approaches to the vacuum value 
gradually with oscillation. 
In Fig.2, the time evolution of the amplitude of quantum sigma meson mode 
with $n=0$, that is $u_0^{{\bf k}={\bf 0}}$, is shown. This mode function 
only oscillates. On the other hand, as is seen in Fig.3, 
the amplitude of the quantum pion mode with $n=0$, 
that is $u_1^{{\bf k}={\bf 0}}$, is amplified. 
The amplitude of pion mode with $n=1$ is also amplified weakly as is 
seen in Fig.4. 
However, as is seen, for example, in Fig. 5, 
the amplification is not realized in the modes with $n \geq 2$. 
Also, the sigma meson modes with $n\ge 1$ as well as $n=0$ are not amplified 
as is seen in Fig.6. 
Thus, we conclude that, in this parameterization, the amplitudes of 
the lowest and the first excited pion modes are amplified, but other 
quantum meson modes are not grown.

\begin{figure}[b]
 \parbox{\halftext}
 {  \epsfxsize=5.5cm  
  \centerline{\epsfbox{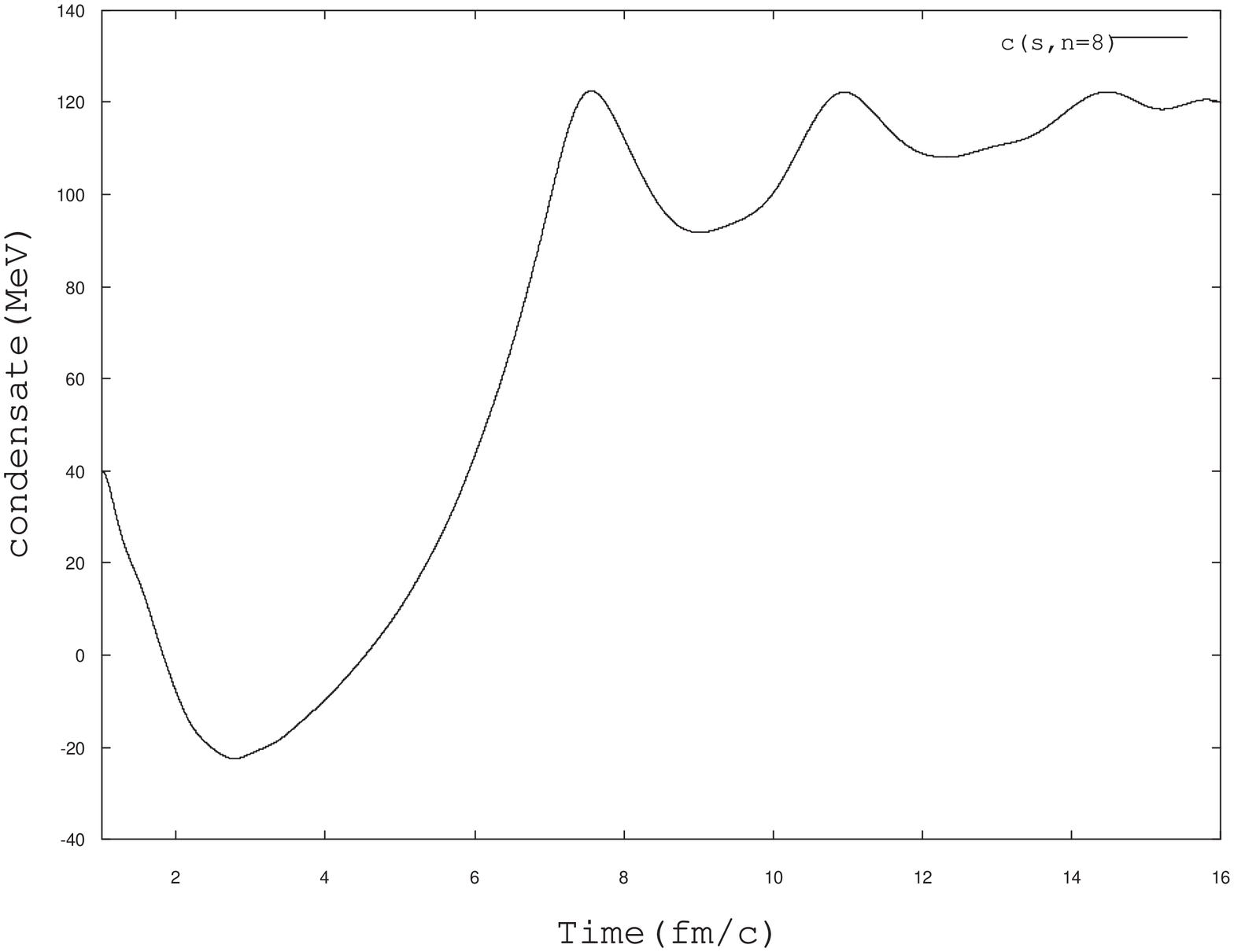}}
  \caption{The time evolution of the chiral condensate with one dimensional 
  expansion is depicted.}}
   \label{fig:bj-condens}
 \hspace{3mm}
 \parbox{\halftext}
 {  \epsfxsize=5.5cm  
  \centerline{\epsfbox{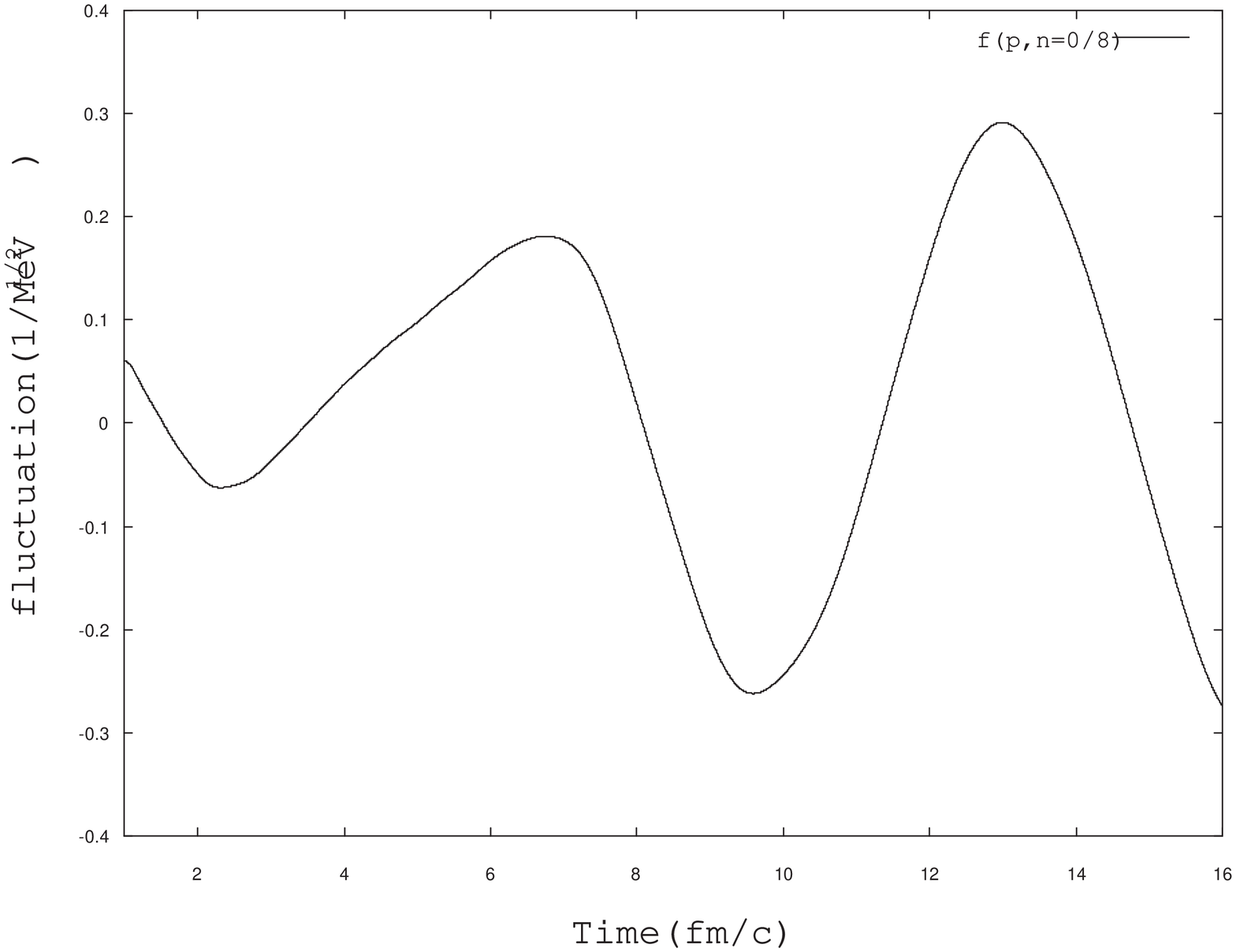}}
  \caption{The time evolution of the fluctuation mode with $n=0$ 
                in the $\pi$-direction is depicted.}}
   \label{fig:bj-pi0}
\end{figure}
\begin{figure}[htb]
 \parbox{\halftext}
 {  \epsfxsize=5.5cm  
  \centerline{\epsfbox{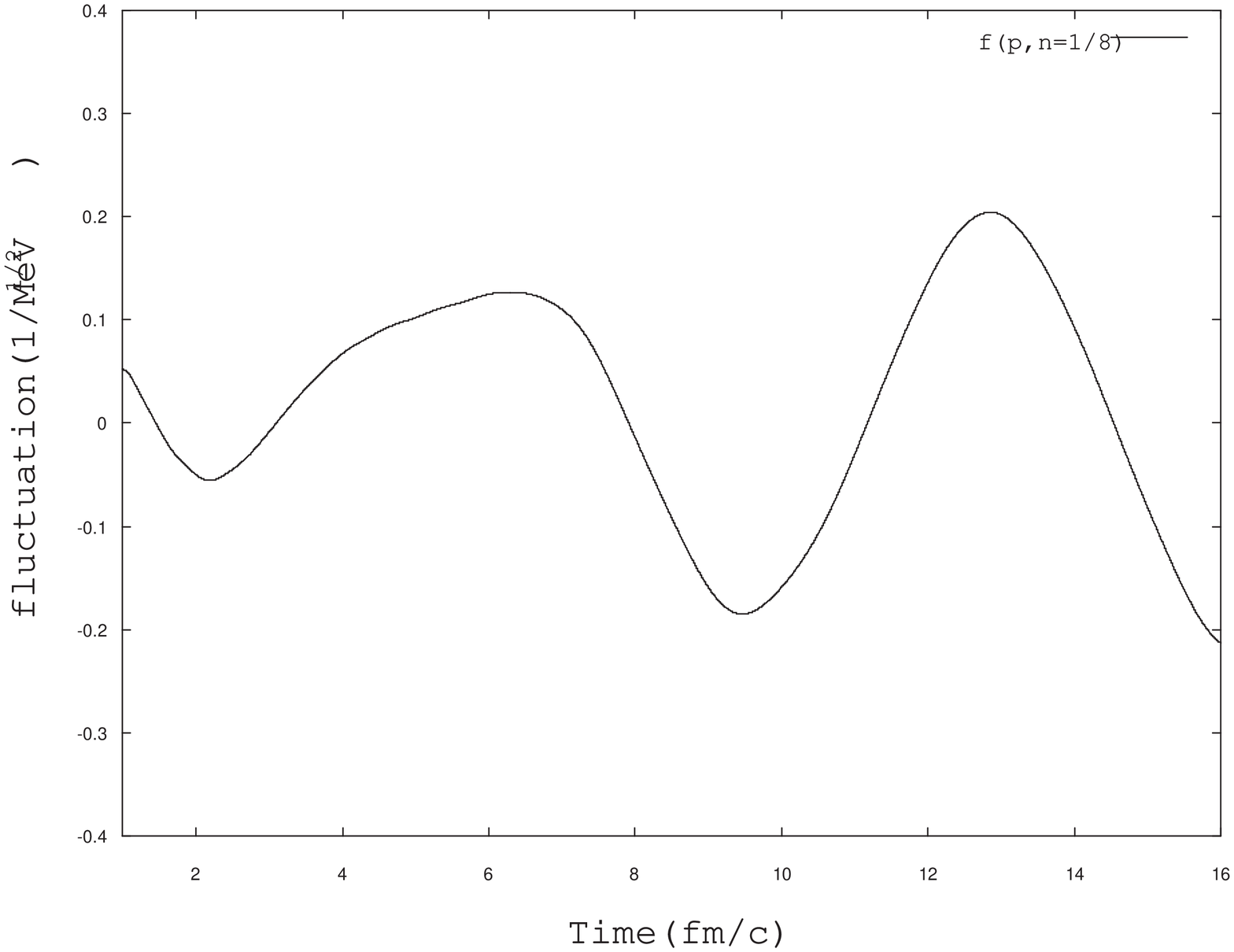}}
  \caption{The time evolution of the fluctuation mode with $n=1$ 
                in the $\pi$-direction is depicted.}}
   \label{fig:bj-pi1}
 \hspace{3mm}
 \parbox{\halftext}
 {  \epsfxsize=5.5cm  
  \centerline{\epsfbox{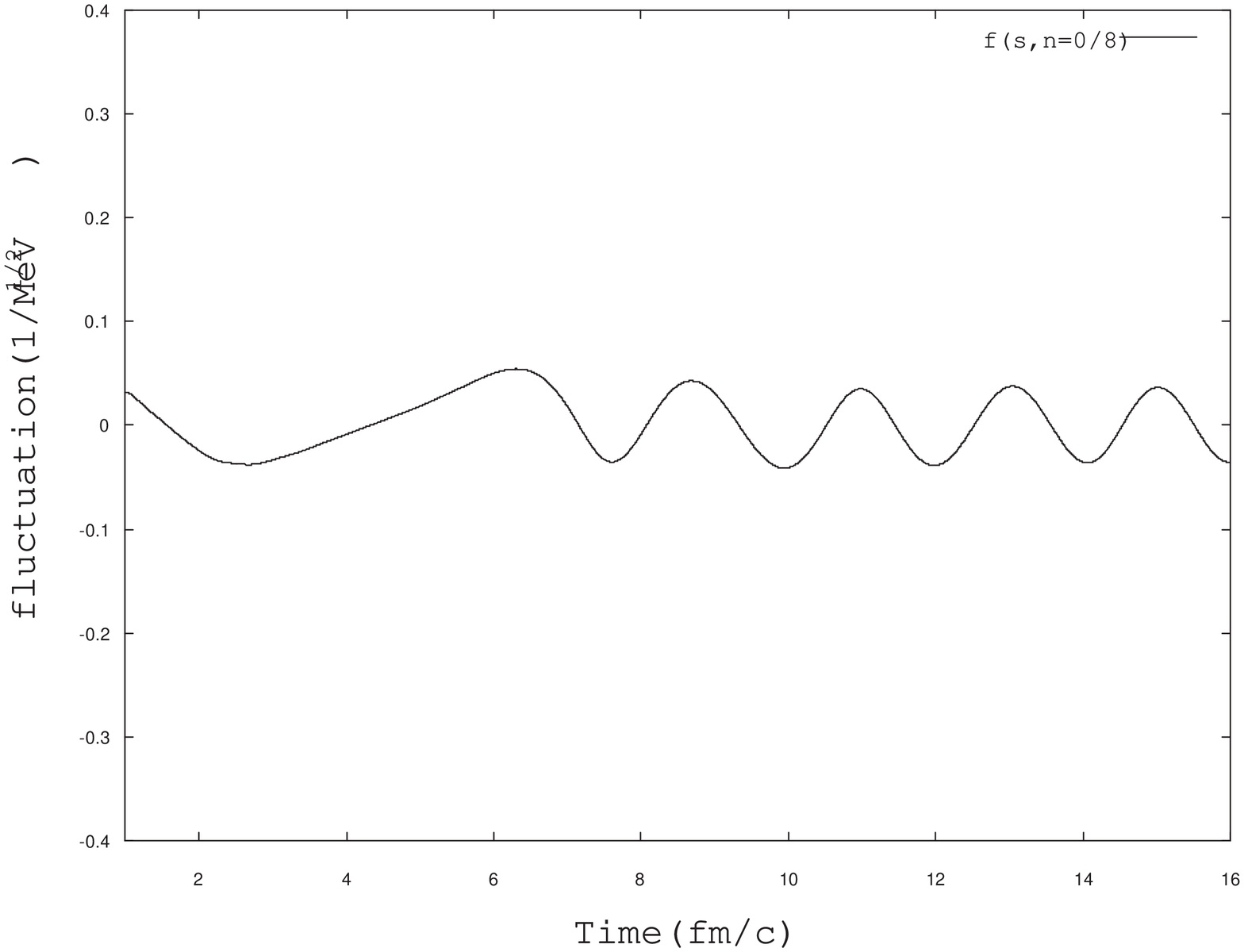}}
  \caption{The time evolution of the fluctuation mode with $n=0$ 
                in the $\sigma$-direction is depicted.}}
   \label{fig:bj-sigma0}
\end{figure}

Next, let us demonstrate qualitatively the time evolution of mean field 
and of quantum meson mode functions in Eqs.(\ref{2-27}) 
and (\ref{2-28}) with spatially expansion.
We assume the same conditions with respect to the chiral condensate as those 
in the case of the homogeneous condensate, 
namely $\bar{\varphi}_0 \neq 0$ and $\bar{\varphi}_i = 0$ for $i=1\sim 3$, 
and the same initial condition for $\tau$ instead of $t$. 
We then impose the periodic boundary conditions for the fluctuation modes,
$k_x = (2\pi/L)n_x,\ k_y = (2\pi/L)n_y $ and 
$k_{\eta} = (2\pi/\alpha)n_{\eta}$,
where $n_x, n_y, n_{\eta}$ are integer, $L$ is spatial length in 
$x$-$y$ direction
and $\alpha$ is dimensionless parameter. 
The fluctuation modes labeled by $(n_x,n_y,n_{\eta})$ are counted up to 
$(\frac{2\pi}{L})^2(n_x^2+n_y^2)
+(\frac{2\pi}{\alpha})^2 (\frac{n_{\eta} + \frac{1}{4}}{\tau})^2 
\leq 1{\rm GeV} $,
where we take $L = 10\ {\rm fm}$ and $\alpha = 4$.
In Fig.7, the time evolution of chiral condensate with the 
expanding geometry is depicted.
It is seen that the behavior of damped oscillation is realized 
quicker than the case without spatially expansion. 
In Figs. 8 and 9, the time evolution of the lowest ($n=0$) and 
the first excited ($n=1$) pion modes is depicted. 
They show that the amplitudes of these modes are amplified weakly. 
However, the sigma modes are not amplified even the lowest sigma mode 
($n=0$) as is seen in Fig. 10.

It should be here noted that the amplification occurs in the late time of 
chiral phase transition both the case of no spatially 
expansion and of the one-dimensional spatial expansion. 
Namely, even when 
the chiral condensate oscillates around its vacuum value with small 
amplitude, there are amplified solutions of pion modes. 
The mechanism of amplification is clarified in the next section.

\section{Late time of chiral phase transition}

In this section, 
we investigate the time evolution of the chiral condensate 
and quantum meson modes based on Eqs.(\ref{2-20}) and (\ref{2-21}) or 
(\ref{2-27}) and (\ref{2-28}) 
in the late time of chiral phase transition. 
Especially, we concern with the amplification of quantum meson modes, 
which was shown in the previous section. 
It is shown that possible mechanism may be an amplification by the forced 
oscillation as well as the parametric resonance.

\subsection{Linear approximation around static configurations 
without expansion}

Hereafter, we assume that fluctuation modes of the pion fields 
are identical one another as was assumed in the previous subsection. 
Also, the condensate depends on time only. 
When the explicit chiral symmetry breaking term, $h$, is small, 
the static solutions in Eq.(\ref{2-20}) and (\ref{2-21}) are given as 
\begin{eqnarray}\label{3-1}
& &\bar{\varphi}_0 = \varphi_0 \equiv 
\sqrt{\frac{3}{\lambda} M_0^2 } - 
\frac{h}{2M_0^2} \ , \nonumber\\
& &u_a^{\bf k}(t) = {u_a^{\bf k}}^s(t) 
\equiv \frac{1}{\sqrt{2E_a^k} } e^{-i E_a^k t} \ ,
\end{eqnarray}
where 
$E_a^{k}=\sqrt{{\mib k}^2 + M_a^2}$. 
The Fourier mode $1/\sqrt{2E_a^k}$ has been determined by the normalization 
condition (\ref{normalization}).

Let us investigate the time-dependent solutions around the above static 
configurations. 
The condensate ${\bar \varphi}_0$ and quantum meson modes $u_a^{\bf k}$ 
can be expanded around the static solutions of Eq.(\ref{3-1}) :
\begin{eqnarray}\label{3-2}
&&{\bar \varphi}_0(t) = \varphi_0 + \delta \varphi(t) \ , \nonumber\\
&&u_a^{\bf k}(t) = {u_a^{\bf k}}^s (t) + \delta u_a^{\bf k}(t) \ . 
\end{eqnarray}
Here, we consider the late time of the chiral phase transition. 
Then, $|\delta \varphi(t)|$ 
is small compared with the vacuum value $\varphi_0$.
Further, we assume that $|\delta u_a^{\bf k}| \ll |{u_a^{\bf k}}^s|$. 
Also, in general, 
the fluctuation $G$ given in (\ref{2-4}) 
is small compared with $\bar{\varphi}_0^2$. 
Since $G$ can be written in the form in (\ref{G}), 
we conclude that the following relation should be satisfied :
\begin{eqnarray}\label{3-3}
& &\bar{\varphi}_0^2 \gg \frac{1}{(2\pi)^3}\int d^3{\mib k}
 \left| {u_a^{\bf k}} \right|^2\ , \nonumber\\
& & \left|\bar{\varphi}_0 \delta \varphi (t)\right|\!  \gg\!
\left| \frac{1}{(2\pi)^3} \!\!\int\!\! d^3{\mib k}
(\! u_a^{\bf k}(t) \delta {u_a^{\bf k}(t)}^* \!\!
+ u_a^{\bf k}(t)^* \delta {u_a^{\bf k}(t)}\! )\right| . \nonumber\\
& &
\end{eqnarray}
Under the above approximation, the equation of motion for 
$\delta \varphi (t)$ 
and $\delta u_a^{\bf k}(t)$ 
are obtained from Eqs.(\ref{2-20}) and (\ref{2-21}). 
For the condensate $\delta\varphi_0(t)$, 
\begin{equation}\label{3-4}
(\partial_t +M_0^2)\delta\varphi(t)=0
\end{equation}
is obtained. 
A solution of (\ref{3-4}) is written as 
\begin{equation}\label{solution}
\delta\varphi(t)=-\delta\sigma \cos (M_0 t+\phi) \ , 
\end{equation}
where $\delta\sigma$ and $\phi$ are constants. 
We rewrite hereafter $a=0$ and $1\sim 3$ into $\sigma$ and ${\pi}$. 
Let us introduce the following dimensionless variables : 
\begin{eqnarray}\label{3-8}
& &\delta \tilde{u}_{a}^{\bf k}(t') = \sqrt{2M_{\sigma}} 
\delta u_{a}^{\bf k}(t)\ , 
\hspace{5mm}
\gamma t' = M_{\sigma} t \ , \nonumber\\
& &\omega_{\sigma}^2 = \frac{ \gamma^2(k^2 + M_{\sigma}^2) }{M_{\sigma}^2}\ , 
\hspace{5mm}
\omega_{\pi}^2 = \frac{ \gamma^2(k^2 + M_{\pi}^2) }{M_{\sigma}^2} \ , 
\nonumber\\
& &h_{\sigma} = \frac{\lambda {\varphi}_0^2}{ k^2 + M_{\sigma}^2 } 
\cdot \frac{\delta \sigma}{{\varphi}_0}\ , 
\hspace{5mm}
h_{\pi} = \frac{1}{3}\cdot\frac{\lambda {\varphi}_0^2}{ k^2 + M_{\pi}^2 } 
\cdot \frac{\delta \sigma}{{\varphi}_0} \ , \nonumber\\
& &F_{\sigma} = \frac{\gamma^2 \lambda {\varphi}_0^2}{ M_{\sigma}^2 } 
\left( \frac{M_{\sigma}^2}{k^2 + M_{\sigma}^2} \right)^{1/4}
\cdot \frac{\delta \sigma}{{\varphi}_0}\ , \nonumber\\
& &F_{\pi} = \frac{\gamma^2 \lambda {\varphi}_0^2}{ 3M_{\sigma}^2 } 
\left( \frac{M_{\pi}^2}{k^2 + M_{\pi}^2} \right)^{1/4}
\cdot \frac{\delta \sigma}{{\varphi}_0}\ , 
\end{eqnarray}
where we put $\gamma=2$. 
If we set up $\phi=0$ without the loss of generality, 
then the equations of motion for the meson modes derived 
from (\ref{2-21}), (\ref{3-1}) and (\ref{3-2}) can simply be expressed as 
\begin{eqnarray}\label{3-10}
\left[ \frac{d^2}{{dt'}^2} 
+ \omega_{\alpha}^2 [1-h_{\alpha}\cos(\gamma t') ] 
\right]\! 
{ \delta {\tilde u}_{\alpha}^{\bf k} }(t') 
= F_{\alpha}\! \cos(\gamma t') e^{-i \omega_{\alpha} t'} \ , 
\end{eqnarray}
where $\alpha=\sigma$ or $\pi$. 
If $F_{\sigma} \hspace{1mm}(F_{\pi})$ is negligible, then 
Eq.(\ref{3-10}) is reduced to Mathieu's equation. 
In this case, the existence of the unstable solution for 
$\delta \tilde{u}_{a}^{\bf k}(t')$ may be expected. 
This phenomena is well known as parametric resonance in classical mechanics.
On the other hand, 
the forced oscillation may be realized by the effect of 
$F_{\sigma} \hspace{1mm}(F_{\pi})$  even if the model parameters do not 
offer the unstable regions for 
the parametric resonance.

\subsection{Linear approximation around static solutions with expansion}

In this subsection, we investigate the case with the expanding geometry.
We derive the equations of motion with expansion using the same formalism in 
\S 4.1 in the late time of chiral phase transition. 
The solution of (\ref{2-28}) is given as 
\begin{eqnarray}\label{3-11}
u_a^{[{\bf k}]}(\tau) = 
{u_a^{[\bf k]}}^s(\tau) \equiv {u_a^{[{\bf k}]}}^s e^{-i E_a^{[{\bf k}]} \tau} 
\ , 
\end{eqnarray}
where $E_a^{[{\bf k}]}=\sqrt{({k_{\eta}^2 + 1/4})/{\tau^2} + k_{\bot}^2 + M_a^2}$. 
We expand the proper-time dependent variables ${\bar \varphi}_0(\tau)$ 
and $u_a^{[{\bf k}]}(\tau)$ in (\ref{2-26}) around ${\varphi}_0$ and 
the above solution (\ref{3-11}), respectively : 
\begin{eqnarray}
&&\bar{\varphi}_0(\tau) = {\varphi}_0 + \delta \varphi_0(\tau) \ , 
\label{3-12}\\
&&u_a^{[{\bf k}]}(\tau)= {u_a^{[{\bf k}]}}^s (\tau) 
+ \delta u_a^{[{\bf k}]}(\tau) \ . 
\end{eqnarray}
Corresponding to (\ref{3-3}), the following approximations are adopted : 
\begin{eqnarray}\label{3-13}
& &\bar{\varphi}_0^2 \gg \frac{1}{\tau}\!\cdot\!
\frac{1}{(2\pi)^3}\!\int\!\!\!\!\int\!\!d^2{k_{\bot}}
\!\!\int\!\!dk_{\eta}
\left| {u_a^{[{\bf k}]}} \right|^2 \ , 
\nonumber\\
& & \bar{\varphi}_0 \delta \varphi_0(\tau) \gg
\frac{1}{\tau}\! \cdot\!
\frac{1}{(2\pi)^3}\!\int\!\!\!\!\int\!\!d^2{k_{\bot}}\!\!\int\!\!dk_{\eta}
  ( {u_a^{[{\bf k}]}}(\tau) \delta {u_a^{[{\bf k}]}}^*\!\!(\tau) 
  + {u_a^{[{\bf k}]}}^*\!\!(\tau) \delta {u_a^{[{\bf k}]}}(\tau)\! ) \ . \ \ 
\end{eqnarray}
Under the above approximation, the equations of motion for 
$\delta \varphi_0(\tau)$ 
and $\delta u_a^{[{\bf k}]}(\tau)$ are obtained from Eqs.(\ref{2-27}) and 
(\ref{2-28}) as
\begin{eqnarray}
&&\left\{ (\partial _{\tau}^2 + \frac{1}{\tau}\partial _{\tau}) 
+ M_\sigma^2(\tau)  \right\}\delta\varphi_0(\tau) = 0 \ , 
 \label{3-14}\\
 && \left( \partial _{\tau}^2 + \frac{ k_{\eta}^2 + 1/4}{\tau^2} + k_{\bot}^2 
 + M_\sigma^2 + \lambda \varphi_0 \cdot \delta\varphi_0(\tau) \right)
 \delta u_\sigma^{[{\bf k}]}(\tau)
 = - \lambda \varphi_0(\tau) \cdot \delta\varphi_0(\tau) 
 u^{[{\bf k}]}_\sigma e^{-i E_\sigma^{[{\bf k}] \tau}} \ , \nonumber\\
 && \left( \partial _{\tau}^2 + \frac{k_{\eta}^2 + 1/4}{\tau^2} + k_{\bot}^2 
 + M_\pi^2 + \frac{\lambda}{3} \varphi_0 \cdot \delta\varphi_0(\tau) \right) 
 \delta u_\pi^{[{\bf k}]}(\tau) 
 =  - \frac{\lambda}{3} \varphi_0(\tau) \cdot \delta\varphi_0(\tau) 
 u^{[{\bf k}]}_\pi e^{-i E_\pi^{[{\bf k}] \tau}} \ .
\nonumber\\
& & \label{3-15}
\end{eqnarray}
Here, 
we have omitted $\partial E_a^{[{\bf k}]}/\partial \tau$ 
because we will replace $\tau$ into $\tau_c$ in the second term 
in (\ref{2-23}) in order to stress the effect of friction due to 
the spatial expansion, where $\tau_c$ may be taken as the proper time 
when the hadronization occurs. Thus, 
by putting $\tau$ into $\tau_c$ of the second term in the left-hand side 
in (\ref{3-14}), 
the amplitude of chiral condensate around the vacuum value reveals 
a behavior of a damped oscillation. 
As is well known, this equation has three-type solutions. For 
$(1/\tau_c)^2 < 4 M_\sigma^2$, we obtain 
\begin{eqnarray}\label{3-16}
\delta {\varphi}_0(\tau) 
= -\delta \sigma e^{-\tau /2\tau_c } \cos(\omega \tau + \phi) \ ,
\end{eqnarray}
where $\delta \sigma$ and $\phi$ are determined by the initial conditions 
and 
\begin{eqnarray}
\omega = \sqrt{M_{\sigma}^2 - \frac{1}{4\tau_c^2}} \ .
\end{eqnarray}
Here, we adopt $\phi = 0$ for simplicity.
We define the following dimensionless variables :\break
\begin{eqnarray}\label{3-17}
& &\delta{\tilde u}_\alpha^{[{\bf k}]}(t')
=\sqrt{2\omega}\delta{u}_\alpha^{[{\bf k}]}(\tau)\ , \qquad
\omega \tau \equiv \gamma t' \ , 
\qquad
\mu = \frac{\gamma}{\sqrt{4M_{\sigma}^2 \tau_c^2 -1}} \ , 
\nonumber\\
& &\omega_{\alpha}^2 = 
\frac{\gamma^2 \left( (k_{\eta}^2 + 1/4)/{\tau_c^2} 
+ k_{\bot}^2 + M_{\alpha}^2 \right)}{\omega^2} \ , 
\nonumber\\
& &h_{\sigma} = 
\frac{\lambda {\varphi}_{0}^2}{({k_{\eta}^2 + 1/4})/{\tau_c^2} 
+ k_{\bot}^2 + M_{\sigma}^2} 
\cdot \frac{\delta \sigma}{{\varphi}_{0}} \ , \qquad
F_{\sigma} = \frac{\gamma^2 \lambda {\varphi}_{0}^2 }{\omega^2}\cdot
 \sqrt{2\omega}u_{\sigma}^{[{\bf k}]} 
\cdot \frac{\delta \sigma}{{\varphi}_{0}} \ , 
\nonumber\\
& &
h_{\pi} = \frac{1}{3}\cdot
\frac{\lambda {\varphi}_{0}^2}{({k_{\eta}^2 + 1/4})/{\tau_c^2} 
+ k_{\bot}^2 + M_{\pi}^2} 
\cdot \frac{\delta \sigma}{{\varphi}_{0}} \ , \qquad
%
%
F_{\pi} = \frac{\gamma^2 \lambda {\varphi}_{0}^2 }{3\omega^2}\cdot
 \sqrt{2\omega}u_{\pi}^{[{\bf k}]} 
\cdot \frac{\delta \sigma}{{\varphi}_{0}} \ , 
\nonumber\\
& &
\end{eqnarray}
where $\alpha=\sigma$ or $\pi$. 
By introducing the above dimensionless variables, 
$\delta{\varphi}_0$ and the equation of motion for quantum meson modes 
in (\ref{3-15}) are recast into 
\begin{eqnarray}
&&\delta \varphi_0(t') 
= -\delta \sigma e^{- \mu t'} \cos(\gamma t') \ , 
\label{3-18}\\
&&\left\{ \frac{d^2}{{dt'}^2} + \omega_{\alpha}^2 [1-h_{\alpha} 
\cos(\gamma t') e^{- \mu t'} ] \right\} 
\delta {\tilde u}_{\alpha}^{[{\bf k}]}(t') 
= F_{\alpha} \cos(\gamma t') e^{- \mu t'} e^{-i \omega_{\alpha} t'} \ . \qquad
\label{3-19}
\end{eqnarray}
The unstable regions in which the amplified solutions of 
(\ref{3-10}) and (\ref{3-19}) exist 
will be investigated in the next section numerically.

%

\section{Unstable regions for quantum meson modes}

\begin{figure}[b]
 {  \epsfxsize=6.0cm  
  \centerline{\epsfbox{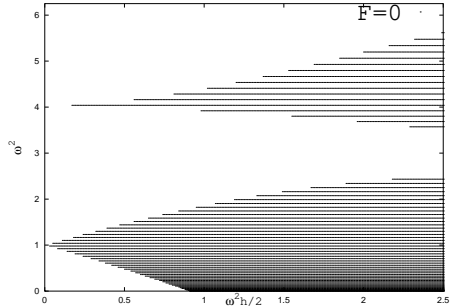}}
                \caption{The unstable regions are depicted in 
                the usual Mathieu equation.
                }}
  \label{fig:fig11}
\end{figure}
\begin{figure}[b]
 \parbox{\halftext}
 {  \epsfxsize=6.0cm  
  \centerline{\epsfbox{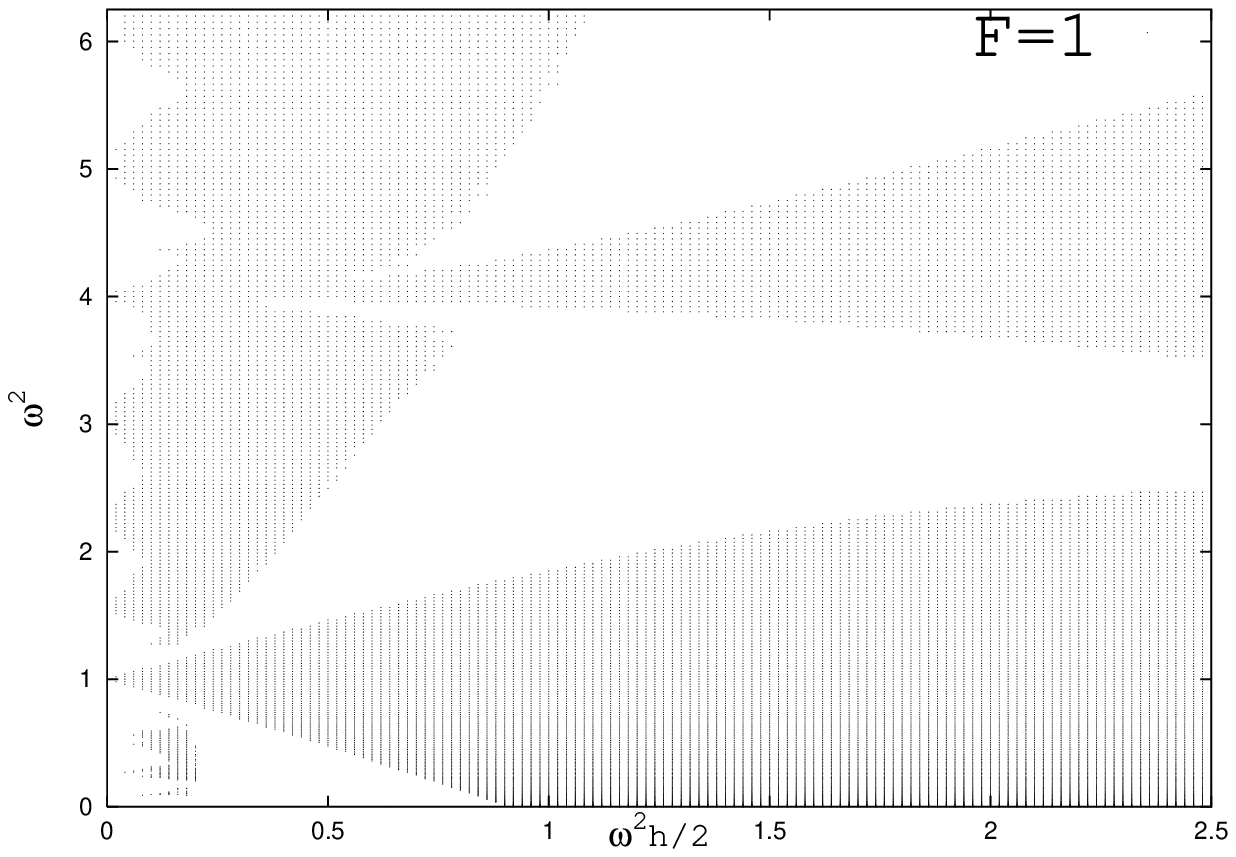}}
                \caption{The unstable regions are depicted for the solutions 
                in Eq.(\ref{3-10}) with $F_{\alpha}=1.0$.}}
 \hspace{3mm}
 \parbox{\halftext}
 {  \epsfxsize=6.0cm  
  \centerline{\epsfbox{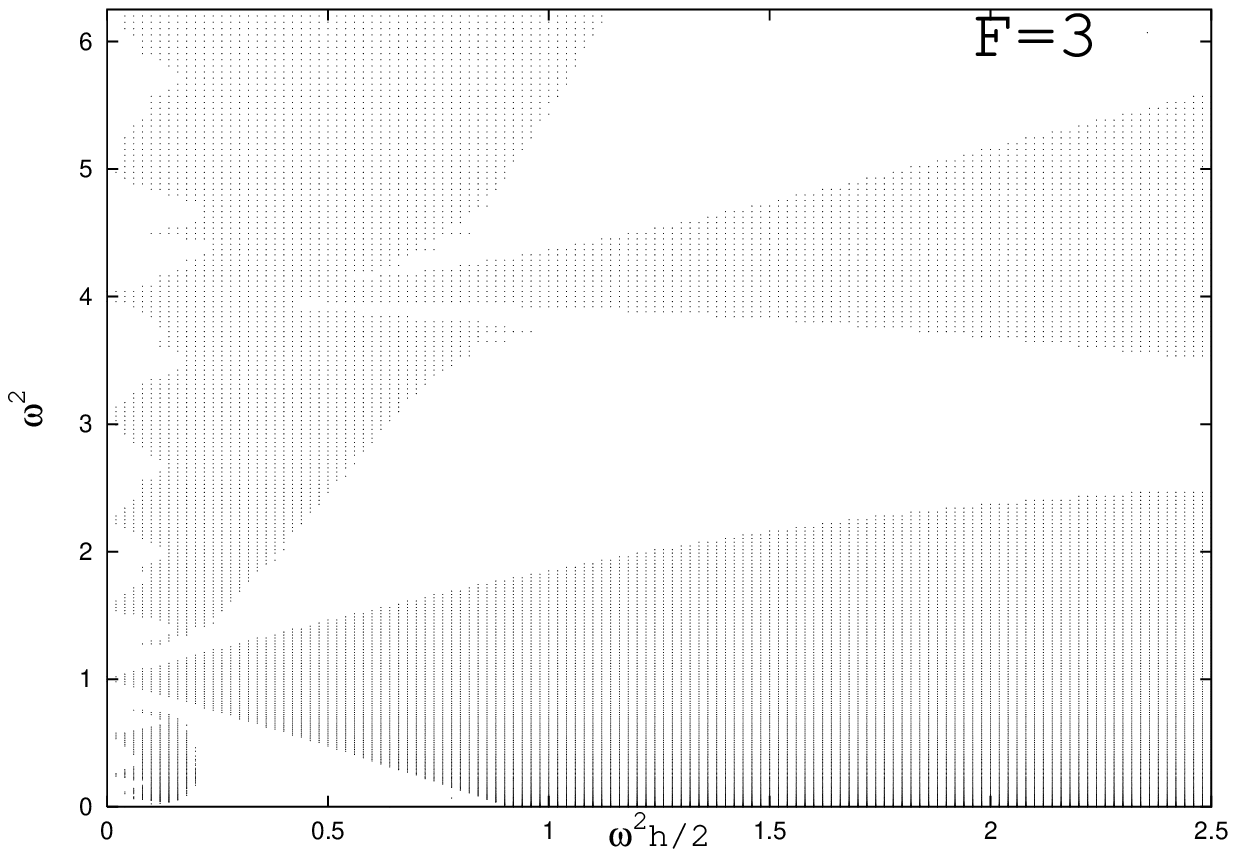}}
                \caption{The unstable regions are depicted for the solutions 
                in Eq.(\ref{3-10}) with $F_{\alpha}=3.0$.}}
\end{figure}

In this section, we numerically 
show the unstable regions in which the absolute 
values of the amplitudes $\delta {\tilde u}$ derived from solutions of  
the equations of motion for quantum meson modes in (\ref{3-10}) without 
spatial expansion and in (\ref{3-19}) with spatially expansion, respectively, 
are amplified. 
This phenomena are seen in the lowest and the first excited pion modes 
as was demonstrated in the previous numerical calculations. 

Hereafter, we omit the subscript $\alpha$. 
In Fig.11, the unstable regions, which are represented in 
the dark area on the figure, for the usual Mathieu equation 
are depicted, which corresponds to the case $F_{\alpha}=0$ in 
(\ref{3-10}). The horizontal axis represents $\omega^2 h/2$ and the vertical 
axis 
represents $\omega^2$, respectively. 
On the other hand, for $F_\alpha\neq 0$, 
the unstable regions are added to those given in the usual Mathieu equation 
because of the effect of the right-hand side 
in (\ref{3-10}) which reveals the effect of the forced oscillation 
or the beat. 
Here and hereafter, 
we obtain the unstable regions, in which the amplification of 
the quantum fluctuation mode functions occurs, as follows : 
We performed the time-integration in a certain time interval of 
the magnitude of the quantum meson mode function, which means a kind 
of the time 
average. Then, we compare a time-integration in a certain time step 
with the proceeding one. 
If the value of time-integration is larger than the last one for all 
time steps under consideration, then we decided that this mode is unstable.
In Figs.12 and 13, the cases $F_{\alpha}=1.0$ and 3.0 
are shown, where the time-integration is taken in 10 fm$/c$ in order to 
judge the unstable or stable solutions up to 100 fm$/c$. 
From these results, it is obviously found that the unstable 
regions become wider due to the effects of the forced oscillation and/or the 
beat induced by the oscillation of the chiral condensate.

\begin{figure}[t]
 {  \epsfxsize=6.0cm  
  \centerline{\epsfbox{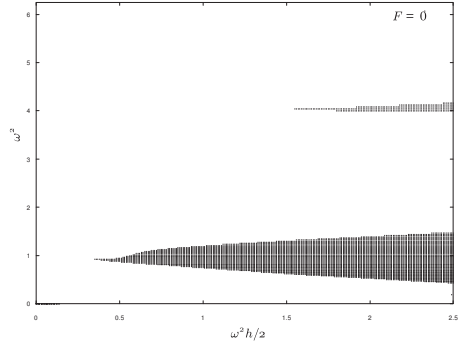}}
                \caption{The unstable regions are depicted in 
                the Mathieu equation with friction term in (\ref{3-19}), 
                in which we set $F_\alpha=0$.}}
\end{figure}
\begin{figure}[t]
 \parbox{\halftext}
 {  \epsfxsize=6.0cm  
  \centerline{\epsfbox{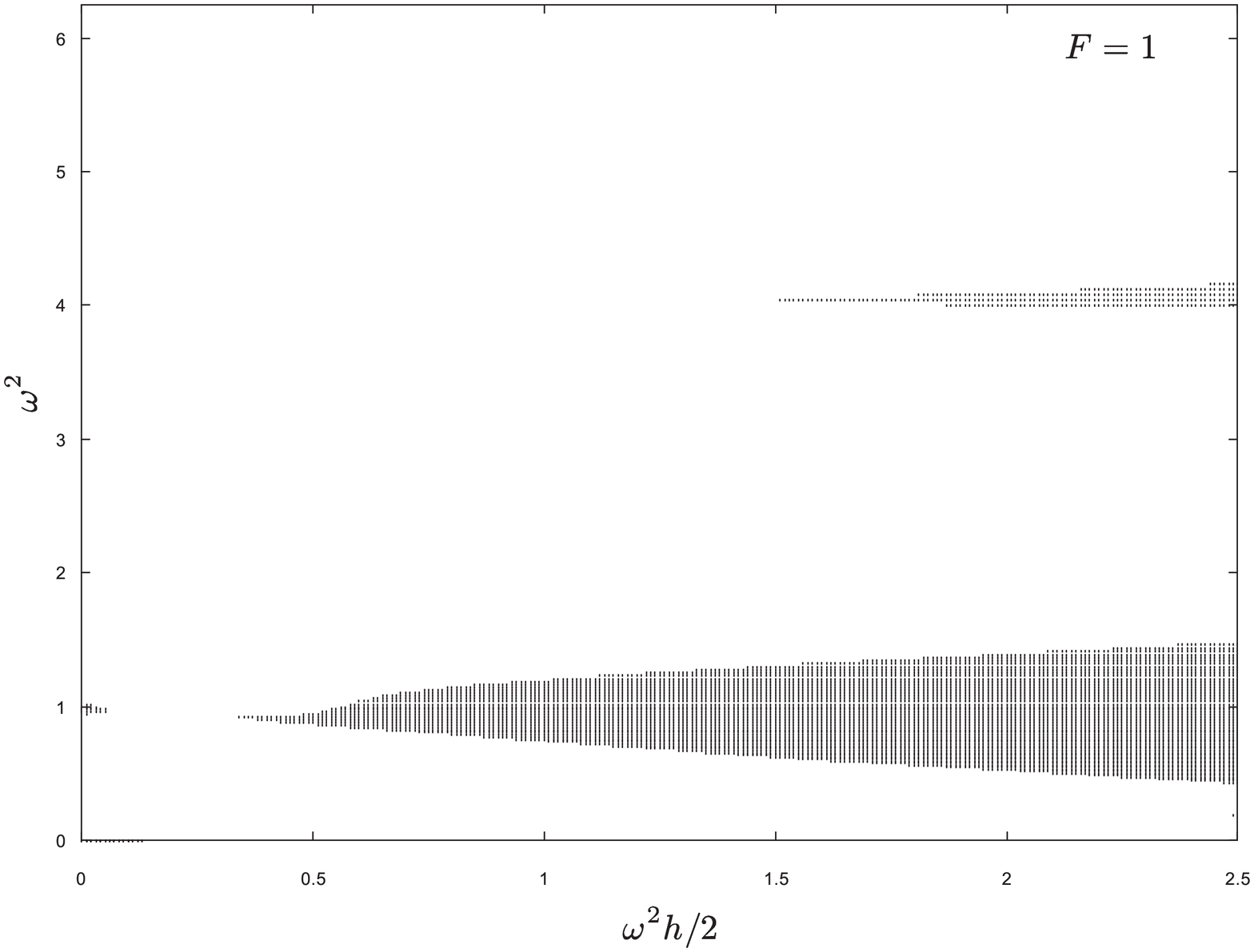}}
                \caption{The unstable regions are depicted for the solutions 
                in Eq.(\ref{3-19}) with $F_{\alpha}=1.0$.}}
 \hspace{5mm}
 \parbox{\halftext}
 {  \epsfxsize=6.0cm  
  \centerline{\epsfbox{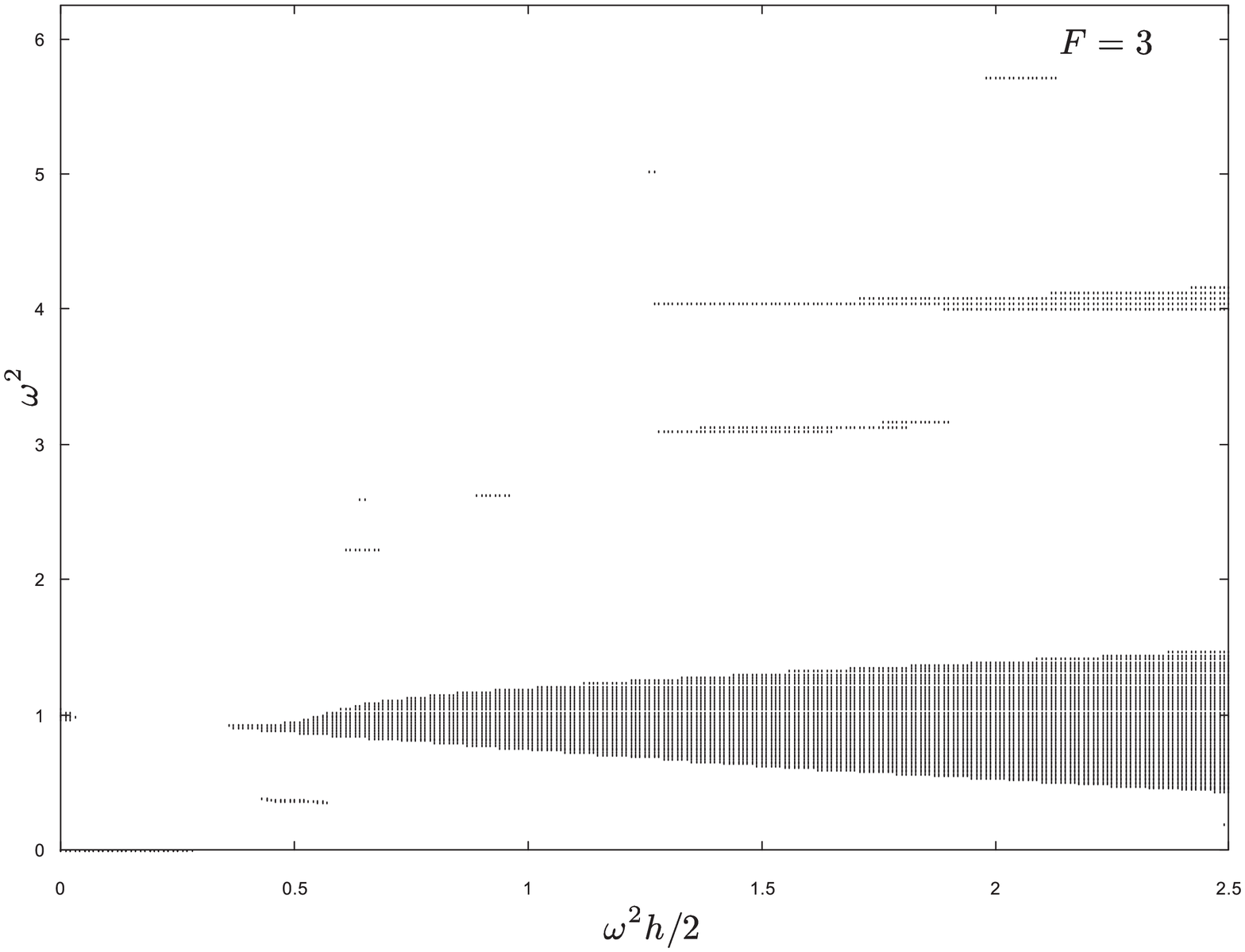}}
                \caption{The unstable regions are depicted for the solutions 
                in Eq.(\ref{3-19}) with $F_{\alpha}=3.0$.}}
\end{figure}

For the case in Eq.(\ref{3-19}) where the friction term appears 
in the equation of motion because the system has the spatially expanding 
geometry, 
the unstable regions are depicted with $F_\alpha=0$, 1 and 3 in 
Figs.14, 15 and 16, respectively. 
The parameter $\mu$ is determined by adopting $M_{\sigma}$ = 500 MeV 
and $\tau_c$ = 10 fm$/c$ .
The time interval for the time-integration to determine the unstable regions 
in our approach is taken as rather short time, 4 fm$/c$,  
because the unstable regions are disappeared 
by strong damping effect for long time past. We calculate the time evolution 
up to 40 fm$/c$. 
In Fig.14 with $F_\alpha=0$, the result for the unstable regions is 
well known because this case corresponds to the case of the 
Mathieu equation with a friction term. 
In Fig. 15 and 16, we introduce the effect of $F_\alpha$ in the realistic 
situation in Eq.(\ref{3-19}) which gives the forced oscillation or the beat. 
However, these effects only give a small modification. 
The reason is that the friction works strongly in this model. 
As a result, the amplification of the quantum meson modes with spatially 
expanding geometry is not so strong, although the amplification appears. 
This results mean that the effect of forced oscillation or the beat 
ceases to work for spatially expanding geometry.


\section{Summary}

We have demonstrated that the amplitudes of quantum pion modes are amplified 
even in the late time of chiral phase transition by the mechanism of 
the forced oscillation as well as the parametric resonance in the 
framework of the $O(4)$ linear sigma model. 
The basic equations of motion have been derived in the time-dependent 
variational approach with the Gaussian wavefunctional, in which 
the equations of motion of the chiral 
condensate and quantum meson mode functions have been formulated 
in the form of the Klein-Gordon type equation and the Liouville-von Neumann 
type equation of motion, respectively, in both the cases of the uniform 
and the one-dimensional spatially expanding systems. 
These equations of motion have been solved numerically, and 
it has been shown that the amplified solutions for the quantum pion modes 
with low momenta exist really. 

We have investigated the mechanism for this amplification phenomena, 
and have pointed out that there is a possibility of the resonance 
by forced oscillation or the beat as well as the parametric resonance 
induced by the oscillation of the chiral condensate around its vacuum value, 
while 
the effect of the forced oscillation ceases to work due to the strong friction 
for the case that has the spatially expanding geometry. 
Also, by numerical calculation, 
we have concretely given the parameter regions 
in which the unstable solutions for the quantum 
meson mode functions exist. 
Of course, it is necessary to know the initial conditions in order to 
judge whether the amplification of quantum meson mode occurs or not, namely, 
the parameters are in the unstable region or not, 
in the realistic relativistic heavy ion collisions.

\section*{Acknowledgements}

K.W. thanks H.Akaike for useful discussions and supports.
Y.T. is supported by the Grants-in-Aid of the Scientific Research 
No.15740156 from the Ministry of Education, Culture, Sports, Science and 
Technology in Japan.


\end{document}